\documentclass[useAMS,usenatbib]{mn2e}

\title[High resolution spectroscopic study of red clump stars in the Galaxy]
      {High resolution spectroscopic study of red clump stars in the Galaxy: 
iron group elements}

\author[E. Puzeras et al.]
    { E. Puzeras,$^{1}$\thanks{E-mail:eduardas.puzeras@tfai.vu.lt} G. Tautvai\v sien\. e,$^{1}$ J. G. Cohen,$^{2}$ 
D. F. Gray,$^{3}$ S. J. Adelman,$^{4}$ \newauthor I. Ilyin$^{5}$ and Y. Chorniy$^{1}$\\
       $^{1}$Institute of Theoretical Physics and Astronomy, Vilnius University, Go\v{s}tauto 
12, Vilnius 01108, Lithuania\\
       $^{2}$Palomar Observatory, Mail Stop 249-17, California Institute of Technology, Pasadena CA 91125, USA\\ 
       $^{3}$Department of Physics \& Astronomy, University of Western Ontario, London ON N6A 3K7, Canada\\
       $^{4}$Department of Physics, The Citadel, Charleston SC 29409-0270, USA\\
       $^{5}$Astrophysikalisches Institut Potsdam, An der Sternwarte 16, Potsdam 14482, Germany }

\begin{document}

\date{Accepted 2010 ..... Received 2010 ......; in original form 2010 ......}

\pagerange{\pageref{firstpage}--\pageref{lastpage}} \pubyear{2010}

\maketitle

\label{firstpage}

\begin{abstract}
The main atmospheric parameters and abundances of the iron group elements
(vanadium, chromium, iron, cobalt and nickel) are determined for 62 red giant 
``clump" stars revealed in the Galactic field by the {\it Hipparcos} orbiting observatory.  
The stars form a homogeneous sample with the mean value of temperature  
$T_{\rm eff}=4750\pm 160$~K, of surface gravity log~$g=2.41\pm0.26$ and the mean value 
of metallicity ${\rm [Fe/H]}=-0.04\pm0.15$~dex. A Gaussian fit to the [Fe/H] distribution produces the mean 
$\langle{\rm [Fe/H]}\rangle= -0.01$ and dispersion $\sigma_{\rm [Fe/H]}=0.08$. 
The near-solar metallicity and small dispersion of $\sigma_{\rm [Fe/H]}$ of 
clump stars of the Galaxy obtained 
in this work confirm the theoretical model of the {\it Hipparcos} clump by 
Girardi \& Salaris (2001).  This suggests 
that nearby clump stars are (in the mean) relatively young objects, reflecting mainly the 
near-solar metallicities developed in the local disk during the last few Gyrs of its history. 
We find iron group element to iron abundance ratios in clump giants to be close to solar. 

\end{abstract}


\section{Introduction}

The clump of core-helium burning stars is a prominent feature in the colour-magnitude 
diagrams of open clusters. Cannon (1970) predicted that the red clump stars should also be 
abundant in the solar neighbourhood. 
Many photometric studies 
have tried with varying success to identify such stars in the Galactic field 
(see Tautvai\v{s}ien\.{e} 1996 for a review), however we had to wait for the {\it Hipparcos} mission. 
The presence of red clump stars in the solar neighbourhood was clearly demonstrated in the HR 
diagrams by Perryman et al.\ (1995). The {\it Hipparcos} catalogue (Perryman et al.\ 1997) contains 
about 600 clump stars 
with parallax error lower than 10\%, and hence an error in absolute magnitude lower than 
0.12 mag. This accuracy limit corresponds to a distance of about 125~pc within which the 
sample of clump stars is complete. Now it is important to investigate their distributions of 
masses, ages, colours, magnitudes and metallicities, which may provide useful constraints to 
chemical evolution models of the local Galactic disk. Moreover, clump stars may be useful 
indicators of ages and distances for stellar clusters and the Local Group galaxies 
(cf. Hatzidimitriou \& Hawkins 1989, Hatzidimitriou 1991, Udalski 1998, Girardi \& Salaris 2001). 

In this paper we report on the primary atmospheric parameters and the abundances of iron group elements 
in the 62 clump stars of the Galactic field obtained from the high-resolution spectra. 
The results are discussed in detail together with results of other studies of the 
clump stars. 
Preliminary results of this study were published by Tautvai\v{s}ien\.{e} et al.\ (2005) and 
Tautvai\v{s}ien\.{e} \& Puzeras (2008).
 
In Fig.~1, we show a HR diagram constructed for the {\it Hipparcos} stars with 
$\sigma_{\pi}/{\pi}< 0.1$  and $\sigma_{B-V}< 0.025$ mag. On the giant branch is a distinct 
red clump at $B-V \approx 1.0, M(H_p)\approx 1.0$ mag. 
The sample of 63 stars investigated in our 
study is indicated by open circles. 
  
\input epsf
\begin{figure}
\epsfxsize=\hsize 
\epsfbox[25 10 170 120]{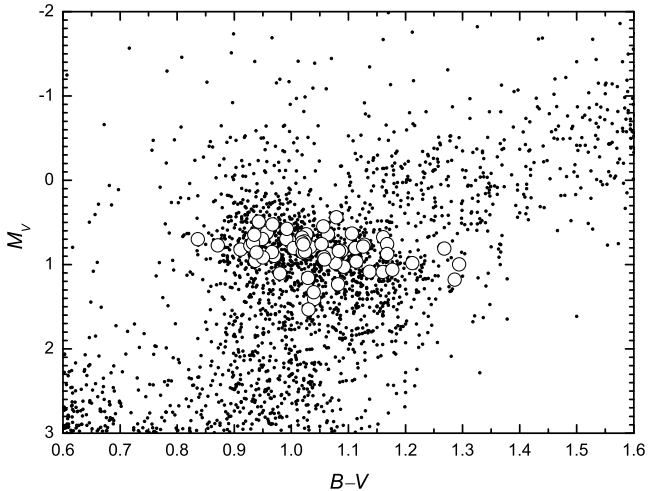} 
\caption{Color-magnitude diagram for the {\it Hipparcos} catalogue.
The programme stars are indicated by open circles.} 
\label{fig1}
\end{figure}

\section{Observational data}

The red clump stars were selected from the {\it Hipparcos} Catalogue (Perryman et al.\ 1997).
Spectra for partially overlapping star samples were observed on several telescopes, described below. 
   
Spectra for 17 stars were obtained at the Nordic Optical Telescope (NOT, La Palma) 
with the SOFIN \'{e}chelle 
spectrograph (Tuominen et al.\ 1999).  
The 2nd optical camera ($R\approx 60\,000$) was used to observe simultaneously 13 spectral orders, 
each of $40-60$~{\AA} in length, located from 5650~{\AA} to 8130~{\AA}. 
Reduction of the CCD images, obtained with SOFIN, was done using the {\em 4A} software package 
(Ilyin 2000). Procedures of bias subtraction, 
cosmic ray removal, flat field 
correction, scattered light subtraction, extraction of spectral orders were 
used for image processing. A Th-Ar comparison spectrum was used for the 
wavelength calibration. The continuum was defined from a number of narrow 
spectral regions, selected to be free of lines.

The spectra of 14 stars were observed with the HIRES spectrograph on the 10-m Keck Telescope.
A 1.1 x 7 arcsec slit ($R\approx 34\,000$) was used, and  
19 spectral orders located from 5620 to 7860~{\AA} were extracted. The 
spectra were reduced using {\it IRAF} 
and {\it MAKEE} packages.

The spectra of 17 stars were observed with the long camera of the 1.22~m Dominion Astrophysical Observatory 
telescope's coud\'{e} spectrograph ($R\approx 40\,000$). The interactive computer graphics program {\it REDUCE} 
by Hill et al.\ (1982) was used to rectify them. The scattered light was removed during the extraction 
procedure by the program {\it CCDSPEC} described in Gullivier \& Hill (2002).

The spectra for 18 stars in the spectral interval from 6220~{\AA} to 6270~{\AA} were obtained at the Elginfield 
Observatory (Canada) with the 1.2~m telescope and the high-resolution coud\'{e} spectrograph ($R\approx 100\,000$). 
Spectra were recorded using an 1872 diode Reticon self-scanned array light detector, mounted in a Schmidt camera, 
with focal length of 559 mm. See Brown et al.\ (2008) for further 
discussion concerning the equipment and operation.

This observational data was supplemented by spectroscopic observations ($R\approx 37\,000$) of red clump stars 
obtained on the 2.16~m telescope of
the Beijing Astronomical Observatory (China) taken from the literature (Zhao et al.\ 2001).

In Fig.~2, we show examples of observed spectra  for several common stars 
using different instruments. 
A careful selection of spectral lines for the analysis has allowed us to avoid 
systematic differences in 
analysis results obtained from the different instruments. 
E.g., the star HD~216228 has been observed on four telescopes, 
a comparison of the measured  equivalent widths (EW) of
its  Fe\,{\sc i} lines is shown in Fig.~3.

\section{Method of analysis and physical data}

The spectra were analysed using a differential model atmosphere technique. 
The programme packages, developed at the Uppsala Astronomical 
Observatory, were used to calculate the theoretical equivalent widths 
and the line profiles. A set of plane parallel, line-blanketed, constant-flux LTE model atmospheres 
was computed with an updated version of the {\it MARCS} code (Gustafsson et al.\ 2003).

The Vienna Atomic Line Data Base (VALD, Piskunov et al.\ 1995) was extensively 
used in preparing the input data for the calculations. Atomic oscillator 
strengths for the spectral lines analysed this study were taken from 
an inverse solar spectrum analysis done in Kiev (Gurtovenko \& Kostik 1989). 

Because of the asymmetric nature of line measurement errors (i.e. problems 
such as blending and telluric line superposition 
always increase measured line width), we used a "quality over quantity" approach when selecting lines for 
abundance calculations. All lines used for calculations were carefully selected.
Inspection of the solar spectrum (Kurucz et al.\ 1984) and the solar line identifications of Moore et al.\ (1966) 
were used to avoid blends and lines blended by telluric absorption lines. All line profiles in all spectra were 
hand-checked  requiring that the line profiles be sufficiently clean to provide reliable equivalent widths. Only 
lines with equivalent widths between 20~m{\AA} and 150~m{\AA} were used for abundance determinations. Spectral 
lines systematically producing outlier abundances in a number of stars, indicating spectral (observational) defect, 
undetected blends or erroneous atomic data, were rejected as well.
The equivalent widths of the lines were measured by fitting of a Gaussian profile using the {\em 4A} software 
package (Ilyin 2000).

Effective temperature, gravity and microturbulence were derived  using  
traditional spectroscopic criteria. The preliminary effective temperatures for 
the stars were determined using the $(B-V)_o$ and $(b-y)_o$ colour indices and the 
temperature calibrations by Alonso et al.\ (2001). For some stars the averaged 
temperatures also include the values obtained from the infrared flux method (IRFM). 
All the effective temperatures were carefully checked and corrected if needed by 
forcing Fe~{\sc i} lines to yield no dependency of iron abundance on excitation 
potential by changing the model effective temperature. Surface gravity was obtained by 
forcing Fe~{\sc i} and Fe~{\sc ii} lines to yield the same [Fe/H] value by 
adjusting
the model gravity. Microturbulence value corresponding to minimal line-to-line  
Fe~{\sc i} abundance scattering was chosen as correct value. Depending upon the 
telescope, the number of Fe~{\sc i} lines analysed was up to 65 and of 
Fe~{\sc ii} up to 12. 

\input epsf
\begin{figure}
\epsfxsize=\hsize 
\epsfbox[15 10 170 135]{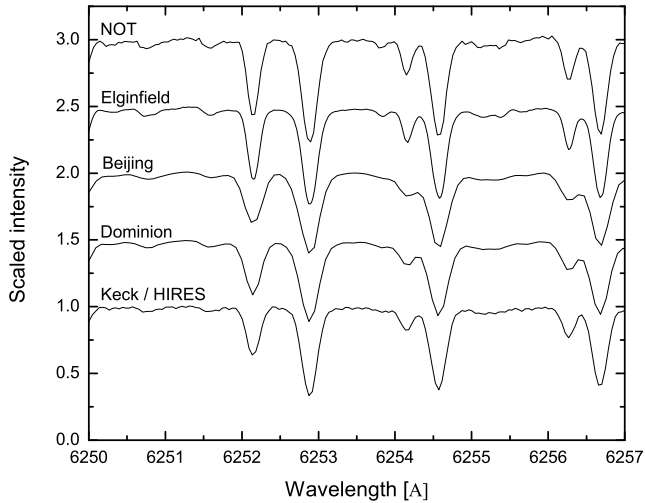} 
\caption{The spectrum of the star HD~216228 in the rgion 
6250 to 6257~\AA\ as observed on four different telescopes.
Since this star was not observed 
on the Keck telescope, we present a spectrum of the very similar star
HD~11037 instead.} 
\label{fig2}
\end{figure}

\input epsf
\begin{figure}
\epsfxsize=\hsize 
\epsfbox[20 10 175 125]{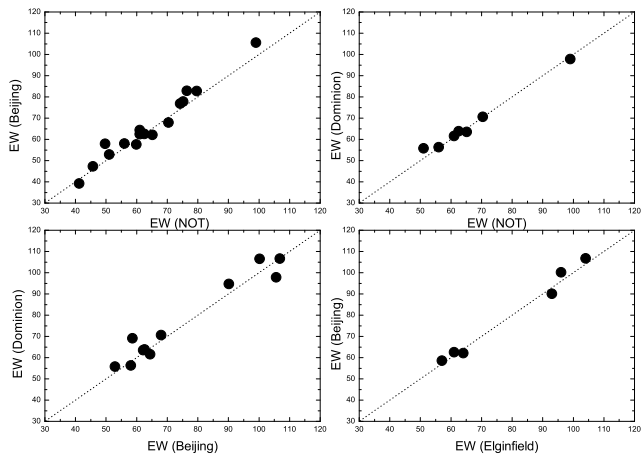} 
\caption{Comparisons of equivalent widths of Fe\,{\sc i} lines in 
overlapping spectral regions observed with different telescopes 
for the star HD~216228. } 
\label{fig3}
\end{figure}

Using the $gf$ values and solar equivalent widths of analysed lines from 
Gurtovenko \& Kostik (1989) we obtained the solar abundances, used later for the 
differential determination of abundances in the programme stars. We used the 
solar model atmosphere from the set calculated in Uppsala with a microturbulent 
velocity of 0.8~$\rm {km~s}^{-1}$, as derived from 
Fe~{\sc i} lines. 

In addition to thermal and microturbulent Doppler broadening of lines, atomic 
line broadening by radiation damping and van der Waals damping were considered 
in the calculation of abundances. Radiation damping parameters of 
lines were taken from the VALD database. 
In most cases the hydrogen pressure damping of metal lines was treated using 
the modern quantum mechanical calculations by Anstee \& O'Mara (1995), 
Barklem \& O'Mara (1997) and Barklem et al.\ (1998). 
When using the Uns\"{o}ld (1955) approximation, correction factors to the classical 
van der Waals damping approximation by widths 
$(\Gamma_6)$ were taken from Simmons \& Blackwell (1982). For all other species a correction factor 
of 2.5 was applied to the classical $\Gamma_6$ $(\Delta {\rm log}C_{6}=+1.0$), 
following M\"{a}ckle et al.\ (1975). For lines stronger than 
$W=100$~m{\AA} the correction factors were selected individually by  
inspection of the solar spectrum. 

Cobalt abundances were investigated with hyperfine structure (HFS) effects taken into account. 
The HFS corrections for every line in every star we calculated with 
the LTE spectral synthesis program {\it MOOG} (Sneden 1973) and the HFS input data adopted from 
Prochaska et al.\ (2000).

\subsection{Estimation of uncertainties}

The sources of uncertainty can be divided into two distinct categories. The first 
category includes the errors that affect a single line (e.g.\ random errors 
in equivalent widths, oscillator strengths), i.e.\ uncertainties of the 
line parameters. Other sources of observational error, such as continuum placement or 
background subtraction problems also are partly included in the equivalent width 
uncertainties. The second category includes the errors which affect all 
the lines together, i.e.\ mainly the model errors (such as errors in the 
effective temperature, surface gravity, microturbulent velocity, etc.). 
The scatter of the deduced line abundances $\sigma$, presented in Table~2 
and Table~3, 
gives an estimate of the uncertainty due to the random errors in 
the line parameters (the mean value of  $\sigma$ is $0.07$).
Thus the uncertainties in the derived abundances that are the 
result of random errors amount to approximately this value. 

Typical internal error estimates for the atmospheric parameters are: 
$\pm 100$~K for $T_{\rm eff}$, $\pm 0.2$~dex for log~$g$ and 
$\pm 0.3~{\rm km~s}^{-1}$ for $v_{\rm t}$. The sensitivity of the abundance 
estimates to changes in the atmospheric parameters by the assumed errors is 
illustrated  for the star HD~218031 in Table~1. It is seen that possible 
parameter errors do not affect the abundances seriously; the element-to-iron 
ratios, for which we use the neutral species for both 
and use in our discussion, are even less sensitive.  

\begin{table}
\centering
\begin{minipage}{80mm}
\caption{The sensitivity of stellar atmosphere abundances to changes in atmospheric parameters. 
Example for the star HD~218031.}
\begin{tabular}{rcccrccc}
\hline
Element & $\Delta T_{\rmn{eff}}$ &  $\Delta$ log g & $\Delta v_{t}$ & $\Delta {\rm Total}$ \\
        &  --100 K                &  +0.2 dex       & +0.3 (km s$^{-1}$) \\
 \hline
   V\,{\sc i}    &--0.17 &  0.00 &--0.08 & 0.19\\  
   Cr\,{\sc i}   &--0.06 &  0.01 &--0.05 & 0.08\\
   Mn\,{\sc i}   &--0.06 &  0.01 &--0.01 & 0.06\\
   Fe\,{\sc i}   &--0.05 &  0.04 &--0.07 & 0.09\\
   Fe\,{\sc ii}  &  0.09 &  0.11 &--0.08 & 0.16\\
   Co\,{\sc i}   &--0.06 &  0.06 &--0.03 & 0.09\\
   Ni\,{\sc i}   &--0.05 &  0.06 &--0.05 & 0.09\\
\hline

\end{tabular}
\end{minipage}
\end{table}

\section{Results and discussion}

Table~2 presents the adopted atmospheric parameters 
for our stellar sample, the numbers of Fe\,{\sc i} and Fe\,{\sc ii} lines 
investigated, the line-to-line scatter and sources of spectral observations used. Some stars 
were observed on several telescopes. For them the main atmospheric parameters 
were not redetermined from the new observations, since several checks gave 
a good agreement. The new observing material was used to increase the number of chemical elements 
with reliable detections. The elemental abundances relative to hydrogen [El/H] of iron group chemical 
elements, line-to-line scatters and numbers of lines investigated are presented in Table~3.

\begin{table*}
\centering
\begin{minipage}{100mm}
\caption{Atmospheric parameters of the programme stars}
\begin{tabular}{rcccrrcccc}
\hline
HD & $T_{\rm eff}$ & log~$g$ & $v_{t}$ & [Fe/H] & $\sigma_{\rm Fe I}$ & ${\rm n}_{\rm Fe I}$ & $\sigma_{\rm Fe II}$  & ${\rm n}_{\rm Fe II}$ & Obs.\\
 \hline
    2910 & 4730 & 2.3 & 1.7 & --0.07 & 0.11 & 52 & 0.08 &  8 & 5\\
    3546 & 4980 & 2.0 & 1.4 & --0.60 & 0.06 & 52 & 0.07 & 12 & 5\\
    3627 & 4360 & 2.1 & 1.8 &   0.01 & 0.12 & 31 & 0.04 &  3 & 2\\
    4188 & 4870 & 2.9 & 1.2 &   0.10 & 0.07 & 11 &  --  &  1 & 4\\
    5268 & 4870 & 1.9 & 1.4 & --0.48 & 0.06 & 47 & 0.06 &  8 & 5\\
    5395 & 4870 & 2.1 & 1.3 & --0.34 & 0.06 & 54 & 0.07 & 10 & 5\\
    5722 & 4910 & 2.3 & 1.3 & --0.14 & 0.05 & 45 & 0.03 &  5 & 2\\
    6805 & 4530 & 2.0 & 1.5 & --0.02 & 0.08 & 39 & 0.05 &  6 & 5\\
    6976 & 4810 & 2.5 & 1.6 & --0.06 & 0.10 & 49 & 0.11 &  8 & 5\\
    7106 & 4700 & 2.4 & 1.3 &   0.02 & 0.07 & 33 & 0.07 &  6 & 1\\
    8207 & 4660 & 2.3 & 1.4 &   0.09 & 0.07 & 31 & 0.04 &  4 & 1\\
    8512 & 4660 & 2.1 & 1.5 & --0.19 & 0.09 & 46 & 0.03 &  6 & 5\\
    8763 & 4660 & 2.2 & 1.4 & --0.01 & 0.06 & 19 & 0.07 &  4 & 1\\
    8949 & 4650 & 2.4 & 1.7 &   0.02 & 0.11 & 51 & 0.10 &  8 & 5\\
    9408 & 4780 & 2.1 & 1.3 & --0.28 & 0.06 & 33 & 0.07 &  6 & 1\\
   11037 & 4830 & 2.3 & 1.2 & --0.02 & 0.07 & 44 & 0.09 &  5 & 2\\
   11559 & 4990 & 2.7 & 1.5 &   0.04 & 0.08 & 51 & 0.03 &  9 & 5\\
   12583 & 4930 & 2.5 & 1.6 &   0.02 & 0.10 & 50 & 0.09 &  8 & 5\\
   15779 & 4810 & 2.3 & 1.2 & --0.03 & 0.05 & 19 & 0.06 &  4 & 1\\
   16400 & 4800 & 2.4 & 1.3 &   0.00 & 0.07 & 34 & 0.04 &  5 & 1\\
   17361 & 4630 & 2.1 & 1.4 &   0.03 & 0.09 & 42 & 0.05 &  6 & 5\\
   18322 & 4660 & 2.5 & 1.4 & --0.04 & 0.07 & 44 & 0.06 &  5 & 5\\
   19476 & 4980 & 3.3 & 1.5 &   0.17 & 0.07 & 38 & 0.08 &  8 & 5\\
   19787 & 4760 & 2.4 & 1.6 &   0.06 & 0.08 & 41 & 0.06 &  7 & 5\\
   25604 & 4770 & 2.5 & 1.6 &   0.02 & 0.08 & 41 & 0.04 &  6 & 5\\
   28292 & 4600 & 2.4 & 1.5 & --0.06 & 0.09 & 40 & 0.05 &  5 & 5\\
   29503 & 4650 & 2.5 & 1.6 & --0.05 & 0.09 & 41 & 0.06 &  5 & 5\\
   34559 & 5060 & 3.0 & 1.5 &   0.07 & 0.07 & 37 & 0.04 &  8 & 5\\
   35369 & 4850 & 2.0 & 1.4 & --0.21 & 0.08 & 49 & 0.04 &  7 & 5\\
   54810 & 4790 & 2.5 & 0.9 & --0.15 & 0.06 & 10 &   -- &  1 & 4\\
   58207 & 4800 & 2.3 & 1.2 & --0.08 & 0.06 & 33 & 0.05 &  6 & 1\\
   61935 & 4800 & 2.4 & 1.2 & --0.02 & 0.06 & 14 & 0.04 &  2 & 1\\
   74442 & 4700 & 2.5 & 1.2 &   0.06 & 0.08 & 10 &   -- &  1 & 4\\
   82741 & 4850 & 2.5 & 1.0 &   0.00 & 0.05 & 12 &   -- &  1 & 4\\
   86513 & 4590 & 2.3 & 1.1 &   0.13 & 0.08 & 16 & 0.05 &  5 & 3\\
   94264 & 4730 & 2.7 & 1.0 & --0.04 & 0.08 & 11 &   -- &  1 & 4\\
   95272 & 4670 & 2.5 & 1.3 &   0.00 & 0.09 & 10 &   -- &  1 & 4\\
  100006 & 4590 & 2.3 & 1.2 & --0.03 & 0.10 & 22 & 0.09 &  5 & 3\\
  104979 & 4880 & 2.1 & 0.9 & --0.24 & 0.06 & 12 &   -- &  1 & 4\\
  108381 & 4700 & 3.0 & 1.4 &   0.25 & 0.09 &  8 &   -- &  1 & 4\\
  131111 & 4740 & 2.5 & 1.1 & --0.17 & 0.05 & 34 & 0.06 &  6 & 1\\
  133165 & 4590 & 2.3 & 1.2 & --0.05 & 0.09 & 19 & 0.08 &  5 & 3\\
  141680 & 4900 & 2.5 & 1.3 & --0.07 & 0.07 & 34 & 0.08 &  5 & 1\\
  146388 & 4700 & 2.5 & 1.3 &   0.18 & 0.06 & 31 & 0.09 &  6 & 1\\
  153210 & 4450 & 2.3 & 1.3 &   0.15 & 0.09 & 54 &   -- &  1 & 3\\
  161096 & 4550 & 2.5 & 1.4 &   0.18 & 0.11 & 64 & 0.07 &  5 & 3\\
  163588 & 4400 & 2.4 & 1.4 & --0.01 & 0.11 & 65 & 0.04 &  5 & 3\\
  169414 & 4550 & 2.3 & 1.5 & --0.09 & 0.09 & 40 & 0.08 &  5 & 2\\
  172169 & 4360 & 2.2 & 1.5 &   0.02 & 0.10 & 40 & 0.07 &  4 & 2\\
  181276 & 4940 & 2.7 & 1.3 &   0.13 & 0.07 & 31 & 0.22 &  2 & 2\\
  188310 & 4730 & 2.5 & 1.2 & --0.11 & 0.07 & 11 &   -- &  1 & 2\\
  188947 & 4760 & 2.5 & 1.3 &   0.06 & 0.08 & 58 &   -- &  1 & 3\\
  197989 & 4760 & 2.3 & 1.1 & --0.07 & 0.07 & 13 &   -- &  1 & 4\\
  203344 & 4730 & 2.4 & 1.2 & --0.06 & 0.06 & 34 & 0.04 &  5 & 1\\
  207134 & 4540 & 2.3 & 1.6 & --0.04 & 0.01 & 41 & 0.07 &  5 & 2\\
  212943 & 4660 & 2.3 & 1.2 & --0.24 & 0.05 & 19 & 0.08 &  4 & 1\\
  216131 & 4980 & 2.7 & 1.3 &   0.07 & 0.07 & 44 & 0.05 &  5 & 2\\
  216228 & 4740 & 2.1 & 1.3 & --0.05 & 0.06 & 15 & 0.07 &  4 & 1\\
  218031 & 4780 & 2.3 & 1.3 & --0.08 & 0.04 & 19 & 0.09 &  4 & 1\\
  219916 & 4980 & 2.6 & 1.4 & --0.04 & 0.08 & 38 & 0.05 &  5 & 2\\
  221115 & 5000 & 2.7 & 1.3 &   0.05 & 0.06 & 18 & 0.04 &  4 & 1\\
  222842 & 4980 & 2.8 & 1.3 & --0.02 & 0.04 & 18 & 0.06 &  4 & 1\\

\hline

\end{tabular}
\medskip
Obs.: 1 -- NOT, 2 -- Keck, 3 -- Dominion, 4 -- Eglifield, 5 -- Beijing. 

\end{minipage}
\end{table*}

The sample of clump stars investigated form quite a homogeneous sample 
with no obvious division into metallicity-dependent groups as was suggested 
by Zhao et al.\ (2001). 
The effective temperature ranges from 4300 
to 5100~K with the mean value $T_{\rm eff}=4750\pm 160$~K;
log~$g$ is between 1.8 and 3.3 with the mean value log~$g=2.41\pm0.26$;
the mean microturbulent velocity $v_{\rm t}=1.34\pm0.19$~km s$^{-1}$;
the metallicity range is from $+0.3$ to $-0.60$~dex, however the majority of stars 
concentrate at the mean value [Fe/H]$=-0.04$~dex with a rms deviation about the mean 
of $0.15$~dex. 

\subsection{Comparison with results by McWilliam (1990)}

There is an overlap of 35 stars between 
our sample of clump stars and that of high-resolution spectroscopic analysis 
by McWilliam (1990). McWilliam \& Rich (1994) noted that the McWilliam (1990) 
study was hampered by the narrow wavelength ($6550-6800$~\AA) coverage and the lack of 
metal-rich model atmospheres, which caused an underestimation of metallicities for metal 
rich stars. Our results confirm this; we find for the stars in common 
[Fe/H]$_{\rm (McW)}-$[Fe/H]$_{\rm (Our)}= -0.13\pm 0.07$.  
      
It is also noticeable that the effective temperature calibrations of McWilliam (1990) and 
of Alonso et al.\ (2001), which was used in our work, have 
a slight systematic temperature determination difference of $50\pm50$~K, 
the temperatures 
of McWilliam being lower.  

\subsection{Comparison with results by Zhao et al. (2001)}

A study of 39 red clump giants was carried out by Zhao et al.\ (2001). 
Unfortunately, the determination of effective temperatures in this work was done 
using the erroneous temperature calibration by Alonso et al.\ (1999), which was 
subsequently updated by Alonso et al.\ (2001). Fig.~4 demonstrates  
the consequences of this misunderstanding on a sample of 24 stars reanalysed in 
our study.  

\begin{figure}
  \begin{center}
\epsfxsize=\hsize 
\epsfbox[20 20 320 220]{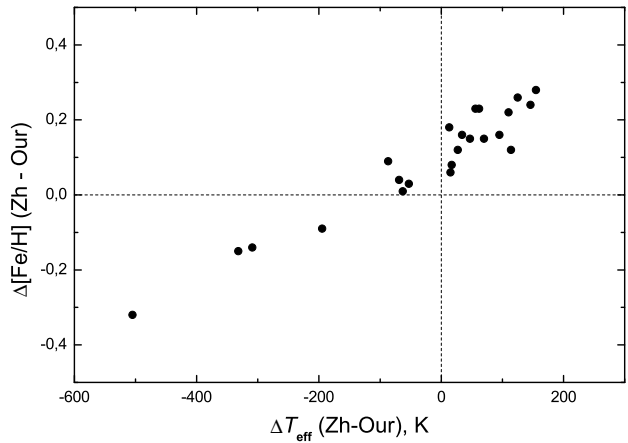} 
  \end{center}
\caption{
The differences of temperature and metallicity caused by the 
the erroneous temperature calibration used by Zhao et al.\ (2001).}     
 \label{fig4}
\end{figure}

For the stars with [Fe/H] greater 
than solar, the temperatures and metallicities were overestimated by up to 
200~K and 0.2~dex, 
respectively. For the metal-deficient stars -- underestimated by approximately the 
same values.
Considering those stars with no temperature difference, it is also 
clear that the [Fe/H] values 
determined in our study are  about 0.1~dex lower. This is
most likely caused  by a very 
careful selection of unblended lines in our analysis and careful choice of 
the continuum level 
in a number of regions. 

\subsection{Comparison with recent studies}

Chemical composition of 177 clump giants of the Galactic disk were investigated 
by Mishenina et al.\ (2006) on a basis of spectra ($R=42\,000$) obtained on the 1.93~m 
telescope of the Haute Provence Observatoire (France).
A sample of 63 Southern hemisphere red clump stars were investigated by Liu et al.\ (2007). 
Spectra ($R=48\,000$) were obtained on the 1.52~m ESO telescope (La Silla, Chile).
A spectroscopic analysis of a sample of 298 nearby giants, with red clump stars among 
them, was done by Luck \& Heiter (2007) based on high resolution spectra ($R=60\,000$) with 
spectral coverage from 4750 to 6850 \AA. We selected a subsample of 138 red clump giants for 
further comparison analysis based on the luminosity and effective temperature diagram provided 
by Luck \& Heiter in their Fig.~20. All stars located in the box limited by 
luminosities log($L/L_{\odot}$) from 1.5 to 
1.8 and effective temperatures from 4700 to 5200~K were included in the subsample. 

A number of stars have been investigated in two or more studies. With one exception, 
iron abundance differences across different papers are negligible and random in nature: 
[Fe/H]$_{\rm Our} - $[Fe/H]$_{\rm LH} = +0.01 \pm 0.06$ (16 stars),  
[Fe/H]$_{\rm Our} - $[Fe/H]$_{\rm L} = +0.04 \pm 0.10$ (8 stars), 
[Fe/H]$_{\rm Our} - $[Fe/H]$_{\rm M} = +0.01 \pm 0.12$ (24 stars), 
[Fe/H]$_{\rm L} - $[Fe/H]$_{\rm LH} = -0.04 \pm 0.04$ (9 stars), 
where Our -- this work, LH -- Luck \& Heiter (2007), M -- Mishenina et al.\ (2006), L -- Liu et al.\ (2007). 
The exception is the clear systematic discrepancy between results of Luck \& Heiter and Mishenina 
et al.: [Fe/H]$_{\rm LH} - $[Fe/H]$_{\rm M} = +0.07 \pm 0.07$ (41 stars). 
It is interesting to note that while the systematic difference of [Fe/H] exists 
between common stars in these two studies, there are no systematic differences between [Fe/H] values in 
our study and studies by Luck \& Heiter and Mishenina et al. Probably this is because the difference 
between these works is more prominent for the metal-deficient 
stars, while our common stars are more metal abundant. So, we can not determine which of these 
studies is right. There are only 3 common stars between Liu et al.\ and Mishenina et al.

Concerning the agreement of effective temperature and surface gravity determinations, our results are in a 
very good agreement with the study by Mishenina et al.\ (2006) despite
the different methods of effective 
temperature determination. So called "spectroscopic" effective temperature values in the work by 
Luck \& Heiter (2007) are systematically higher by almost 100~K, and log~$g$ by +0.4~dex, however 
their "physical" determinations are in a good agreement with our work and Mishenina et al.
Effective temperatures in the work by Liu et al.\ are in the mean by about 60~K lower and log~$g$ by about 
0.3~dex higher than in our work, as evaluated from 8 common stars. The systematic difference in temperatures 
between the studies of Luck \& Heiter (spectroscopic) and Liu et al.\ is 160~K.    

\begin{table*}
\centering
\begin{minipage}{140mm}

\caption{Element abundances of the programme stars}
\begin{tabular}{rrrrrrrrrrrrr}   
\hline
HD & [V/H] & $\sigma$ & n & [Cr/H] & $\sigma$ & n & [Co/H] & $\sigma$ & n & [Ni/H] & $\sigma$ & n \\
 \hline
    2910 &   0.04 & 0.10 & 11 & --0.10 & 0.12 &  4  & --0.01 & 0.08 & 4 &   0.03 & 0.09 &  32 \\
    3546 & --0.46 & 0.06 & 12 & --0.65 & 0.07 &  4  & --0.47 & 0.07 & 4 & --0.47 & 0.07 &  29 \\
    3627 &   0.23 & 0.09 &  6 &   0.07 & 0.11 &  3  &   0.11 & 0.08 & 2 &   0.04 & 0.08 &  10 \\
    4188 &   0.20 & 0.07 &  6 &      - &    - &  -  &   0.15 &    - & 1 &   0.18 & 0.05 &   2 \\
    5268 & --0.53 & 0.10 &  8 & --0.55 & 0.15 &  2  & --0.56 & 0.07 & 4 & --0.48 & 0.10 &  29 \\
    5395 & --0.23 & 0.04 & 12 & --0.26 & 0.06 &  6  & --0.28 & 0.09 & 4 & --0.22 & 0.07 &  26 \\
    5722 & --0.17 & 0.07 &  6 & --0.21 & 0.05 &  5  & --0.26 & 0.03 & 3 & --0.18 & 0.06 &  22 \\
    6805 &   0.19 & 0.08 &  8 & --0.03 & 0.08 &  5  &   0.00 & 0.01 & 3 &   0.13 & 0.11 &  29 \\
    6976 &   0.07 & 0.07 &  9 & --0.11 & 0.05 &  4  &   0.05 & 0.04 & 2 &   0.04 & 0.09 &  31 \\
    7106 &   0.25 & 0.11 & 13 &   0.09 & 0.07 &  6  &   0.13 & 0.08 & 4 &   0.08 & 0.06 &  15 \\
    8207 &   0.17 & 0.09 & 10 &   0.18 & 0.07 &  4  &   0.04 & 0.10 & 3 &   0.18 & 0.08 &  16 \\
    8512 & --0.02 & 0.06 & 10 & --0.23 & 0.06 &  6  & --0.06 & 0.05 & 2 & --0.03 & 0.08 &  25 \\
    8763 &   0.17 & 0.08 & 10 &   0.04 & 0.04 &  2  & --0.06 & 0.07 & 2 &   0.05 & 0.04 &   9 \\
    8949 &   0.18 & 0.06 &  6 & --0.16 & 0.07 &  4  &   0.14 & 0.05 & 2 &   0.18 & 0.09 &  30 \\
    9408 & --0.21 & 0.06 & 13 & --0.33 & 0.02 &  3  & --0.18 & 0.01 & 2 & --0.26 & 0.05 &  17 \\
   11037 & --0.16 & 0.07 &  6 & --0.15 & 0.02 &  4  & --0.14 & 0.00 & 3 & --0.09 & 0.08 &  12 \\
   11559 &   0.13 & 0.06 & 12 &   0.06 & 0.08 &  5  &   0.08 & 0.05 & 5 &   0.12 & 0.08 &  29 \\
   12583 &   0.03 & 0.07 & 10 & --0.02 & 0.09 &  4  & --0.06 & 0.04 & 3 &   0.05 & 0.08 &  30 \\
   15779 &   0.05 & 0.08 & 10 &   0.02 & 0.02 &  2  & --0.13 & 0.06 & 2 & --0.01 & 0.05 &  10 \\
   16400 &   0.01 & 0.06 & 13 &   0.04 & 0.05 &  5  & --0.07 & 0.09 & 3 &   0.02 & 0.07 &  17 \\
   17361 &   0.24 & 0.03 &  5 &   0.08 & 0.07 &  4  &   0.03 &    - & 1 &   0.15 & 0.10 &  32 \\
   18322 &   0.13 & 0.07 &  6 & --0.10 & 0.04 &  5  &   0.02 & 0.06 & 3 &   0.05 & 0.10 &  30 \\
   19476 &   0.33 & 0.05 &  7 &   0.24 & 0.08 &  4  &   0.17 & 0.07 & 6 &   0.28 & 0.08 &  20 \\
   19787 &   0.08 & 0.10 & 11 &   0.06 & 0.11 &  5  &   0.03 & 0.06 & 4 &   0.18 & 0.09 &  30 \\
   25604 &   0.10 & 0.09 & 11 &   0.03 & 0.05 &  5  &   0.12 & 0.09 & 5 &   0.16 & 0.09 &  32 \\
   28292 &   0.14 & 0.09 &  9 & --0.11 & 0.09 &  5  &   0.06 & 0.06 & 2 &   0.07 & 0.10 &  32 \\
   29503 &   0.29 & 0.08 &  7 & --0.06 & 0.10 &  5  &   0.15 & 0.03 & 3 &   0.12 & 0.11 &  25 \\
   34559 &   0.12 & 0.05 & 10 &   0.15 & 0.08 &  5  &   0.03 & 0.06 & 5 &   0.15 & 0.08 &  32 \\
   35369 & --0.23 & 0.05 &  7 & --0.20 & 0.10 &  4  & --0.31 & 0.08 & 4 & --0.13 & 0.07 &  32 \\
   54810 &   0.01 & 0.07 &  5 &      - &    - &  -  & --0.05 &    - & 1 & --0.13 & 0.00 &   2 \\
   58207 &   0.01 & 0.07 & 13 & --0.02 & 0.04 &  6  & --0.08 & 0.08 & 3 & --0.07 & 0.05 &  17 \\
   61935 &   0.00 & 0.05 &  4 &   0.08 & 0.07 &  3  &      - &    - & - & --0.03 & 0.04 &   7 \\
   74442 &   0.25 & 0.12 &  5 &      - &    - &  -  &   0.28 &    - & 1 &   0.20 & 0.08 &   2 \\
   82741 &   0.02 & 0.07 &  6 &      - &    - &  -  &   0.06 &    - & 1 & --0.04 & 0.03 &   2 \\
   86513 &   0.06 & 0.04 & 10 &      - &    - &  -  &      - &    - & - &   0.19 & 0.05 &   9 \\
   94264 &   0.20 & 0.09 &  6 &      - &    - &  -  &   0.05 &    - & 1 &   0.05 & 0.11 &   2 \\
   95272 &   0.14 & 0.10 &  6 &      - &    - &  -  &   0.14 &    - & 1 &   0.13 & 0.14 &   2 \\
  100006 & --0.16 & 0.06 &  8 &      - &    - &  -  &      - &    - & - &   0.00 & 0.06 &  10 \\
  104979 & --0.33 & 0.07 &  7 &      - &    - &  -  & --0.11 &    - & 1 & --0.23 & 0.09 &   2 \\
  108381 &   0.56 & 0.08 &  5 &      - &    - &  -  &   0.50 &    - & 1 &   0.39 & 0.18 &   2 \\
  131111 &   0.01 & 0.07 & 11 & --0.11 & 0.07 &  5  & --0.20 & 0.08 & 3 & --0.18 & 0.05 &  16 \\
  133165 & --0.09 & 0.06 &  7 &      - &    - &  -  &      - &    - & - & --0.04 & 0.04 &   9 \\
  141680 &   0.04 & 0.06 & 13 &   0.02 & 0.08 &  5  & --0.02 & 0.04 & 4 & --0.07 & 0.05 &  16 \\
  146388 &   0.40 & 0.09 & 11 &   0.27 & 0.06 &  5  &   0.14 & 0.08 & 3 &   0.26 & 0.09 &  17 \\
  153210 &   0.13 & 0.13 &  9 &   0.07 & 0.09 &  7  &   0.08 & 0.05 & 3 &   0.24 & 0.12 &  12 \\
  161096 &   0.37 & 0.12 &  7 &   0.22 & 0.09 &  6  &   0.36 & 0.08 & 3 &   0.32 & 0.15 &  12 \\
  163588 &   0.03 & 0.12 &  9 & --0.11 & 0.18 &  7  &   0.01 & 0.07 & 3 &   0.06 & 0.12 &  15 \\
  169414 &   0.20 & 0.06 &  5 & --0.13 & 0.10 &  4  &   0.01 & 0.02 & 2 & --0.11 & 0.09 &  12 \\
  172169 &   0.38 & 0.09 &  6 & --0.08 & 0.07 &  4  &   0.00 &    - & 1 &   0.04 & 0.09 &  10 \\
  181276 &   0.03 & 0.10 &  8 &   0.09 & 0.11 &  4  & --0.08 & 0.04 & 3 &   0.06 & 0.11 & 12 \\
  188310 &   0.10 & 0.08 &  6 &      - &    - &  -  &      - &    - & - &   0.04 & 0.07 &  2  \\
  188947 &   0.29 & 0.05 & 17 &   0.18 & 0.06 &  6  & --0.15 & 0.03 & 2 &   0.19 & 0.08 &  18 \\
  197989 &   0.02 & 0.08 &  6 &      - &    - &  -  & --0.10 &    - & 1 & --0.01 & 0.04 &  3  \\
  203344 &   0.06 & 0.08 & 12 & --0.03 & 0.13 &  3  & --0.02 & 0.13 & 3 & --0.03 & 0.06 &  17 \\
  207134 &   0.26 & 0.08 &  6 & --0.01 & 0.12 &  4  &   0.05 & 0.04 & 3 & --0.04 & 0.09 &  12 \\
  212943 & --0.06 & 0.08 & 10 & --0.21 &    - &  1  & --0.23 & 0.08 & 2 & --0.22 & 0.04 &  9  \\
  216131 & --0.02 & 0.07 &  7 & --0.01 & 0.06 &  5  & --0.14 & 0.02 & 3 & --0.03 & 0.07 &  12 \\
  216228 &   0.08 & 0.07 &  7 &   0.00 & 0.04 &  2  & --0.07 & 0.11 & 2 &   0.00 & 0.03 &  8  \\
  218031 &   0.09 & 0.06 & 10 & --0.07 & 0.05 &  2  & --0.11 & 0.08 & 2 & --0.06 & 0.04 &  10 \\
  219916 & --0.08 & 0.08 &  6 & --0.11 & 0.06 &  5  & --0.20 & 0.06 & 3 & --0.12 & 0.05 &  12  \\
  221115 &   0.09 & 0.05 & 10 &   0.08 & 0.05 &  2  &   0.00 & 0.06 & 2 &   0.06 & 0.04 &  8  \\
  222842 &   0.09 & 0.05 & 10 &   0.06 & 0.05 &  2  & --0.06 & 0.05 & 2 & --0.01 & 0.05 &  9  \\

\hline

\end{tabular}
\end{minipage}
\end{table*}

\section{Metallicity distribution in clump stars of the Galaxy}

There have been few attempts to derive typical metallicities for {\it Hipparcos} clump stars. 
The first was an indirect method by Jimenez et al.\ (1998), they obtained 
$-0.7 < {\rm [Fe/H]} < 0.0$. However, they modelled 
the clump with star formation rate (SFR) strongly decreasing with Galactic age. 
Consequently, they were considering, essentially, the behaviour of the old clump stars, 
with masses of about 0.8 -- 1.4~$M_\odot$. Intermediate-mass clump stars with mass more than 
1.7~$M_\odot$ were absent in their simulations (c.f. Girardi \& Salaris 2001). 
Thus their description of the clump stars was incomplete. 

\input epsf
\begin{figure}
\epsfxsize=\hsize 
\epsfbox[15 5 175 120]{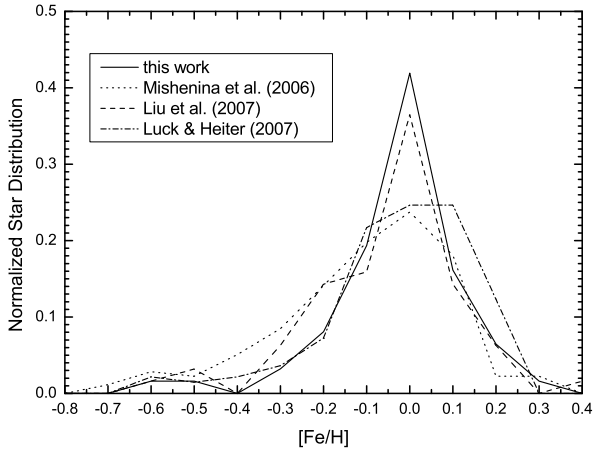} 
 
\caption{Distributions of [Fe/H] in the Galactic clump stars investigated in this work and 
other studies.} 
\label{fig5}
\end{figure}

\input epsf
\begin{figure}
\epsfxsize=\hsize 
 
\epsfbox[15 5 175 120]{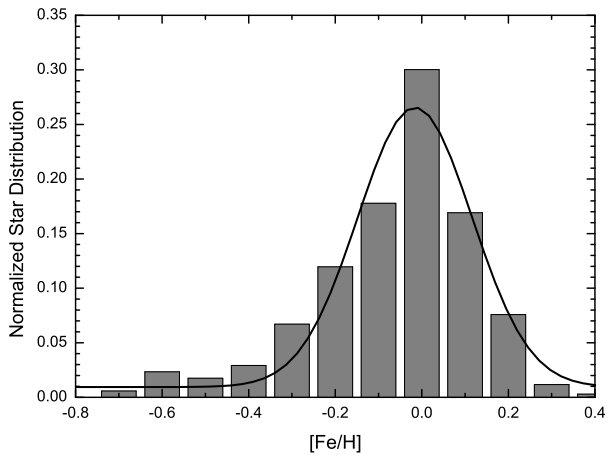} 
\caption{Metallicity distribution in the sample of 342 Galactic clump stars (the [Fe/H] values are 
averaged for the common stars investigated in this study, Mishenina et al.\ (2006), 
Liu et al.\ (2007) and Luck \& Heiter (2007).} 
\label{fig6}
\end{figure}

Girardi et al.\ (1998) considered the full mass range of clump stars, and models with 
constant SFR up to 10 Gyr ago. They demonstrated that the best fit is achieved with a Galaxy 
model which in the mean has solar metallicity, with a very small metallicity dispersion of about 0.1~dex. 
       
Girardi \& Salaris (2001) collected the spectroscopic abundance determinations for the 
{\it Hipparcos} clump stars in the catalogue by Cayrel de Strobel et al.\ (1997). They found 
that the histogram of [Fe/H] values is fairly well represented by a quite narrow Gaussian curve 
of mean $\langle{\rm [Fe/H]}\rangle=-0.12$~dex and standard deviation of 0.18~dex, as derived by means of 
a least-squares fit. 

In the same paper a theoretical simulation of the {\it Hipparcos} clump 
was made. Girardi \& Salaris (2001) found that the total range of metallicities allowed by their model is quite large 
($-0.7 \leq {\rm [Fe/H]} \leq 0.3$), however the distribution for clump stars is very narrow: 
a Gaussian fit to the [Fe/H] distribution produces a mean $\langle{\rm [Fe/H]}\rangle= +0.03$~dex and dispersion 
$\sigma_{\rm [Fe/H]}=0.17$. Actually, the [Fe/H] distribution presents an asymmetric tail at 
lower metallicities, which causes the straight mean of [Fe/H] to be $-0.04$~dex.  
The near-solar metallicity and so small $\sigma_{\rm [Fe/H]}$ imply that nearby clump stars are (in the mean) 
relatively young objects, reflecting mainly the near-solar metallicities developed in the local disk during the 
last few Gyrs of its history. From the 
same simulation they determined that the peak of age distribution in the {\it Hipparcos} clump is at about 1~Gyr. 

In our study, the results of [Fe/H] determinations in clump stars of the Galaxy confirm the theoretical model 
by Girardi \& Salaris (2001). The metallicity range in our study is from 
$+0.3$ to $-0.6$~dex, however the majority of stars concentrate near the mean value 
$\langle{\rm [Fe/H]}\rangle =-0.04\pm0.15$. A Gaussian fit to the [Fe/H] distribution produces the mean 
$\langle{\rm [Fe/H]}\rangle= -0.01$ and very small dispersion $\sigma_{\rm [Fe/H]}=0.08$. 

In order to see what the metallicity distribution is in all the sample of Galactic clump stars investigated to date  
using high resolution spectra, we present in Fig.~5 
metallicity distributions for the samples of Galactic clump 
stars investigated in this study (62 stars), by Mishenina et al.\ (177 stars), Liu et al.\ (63 stars) and 
Luck \& Heiter (138 stars); and in Fig.~6, the metallicity distribution in 
the entire sample 
of 342 Galactic clump stars is presented. The metallicity values were averaged for the stars with multiple analyses. 
In the study of Mishenina et al.\ a special attempt was made to include the metal-deficient stars, so the 
distribution slightly reflects this selection effect.

The [Fe/H] values of clump stars in Fig.~6 range from +0.4 to $-0.8$~dex.
A Gaussian fit to the [Fe/H] distribution produces a mean $\langle{\rm [Fe/H]}\rangle= -0.02$ and dispersion 
$\sigma_{\rm [Fe/H]}=0.13$~dex, which 
is in agreement with the theoretical model by Girardi \& Salaris (2001).

\input epsf
\begin{figure}
\epsfxsize=\hsize 
\epsfbox[5 10 180 245]{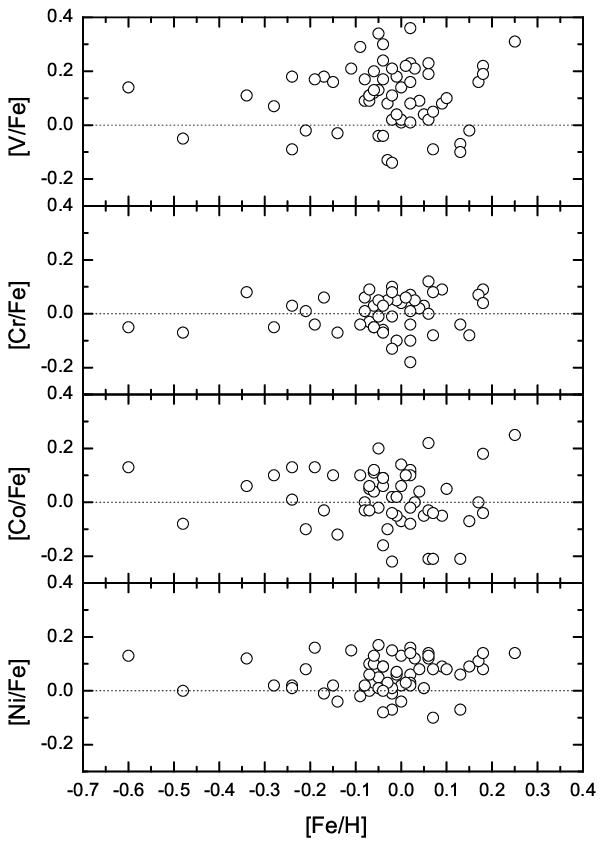} 
\caption{Abundance trends of iron group elements.} 
\label{fig7}
\end{figure}

\subsection{Abundances of iron group elements}

Fig.~7 presents the observed [El/Fe] ratios for iron group elements in our 
sample of stars, always using the neutral species. 
Nickel abundances always closely follow solar nickel to iron ratios in the Galactic disk. 
In our study ${\rm [Ni/Fe]} = 0.06\pm 0.07$, in Mishenina et al.\ (2006) -- ${\rm [Ni/Fe]} = 0.11\pm 0.03$, 
in Liu et al.\ (2007) -- ${\rm [Ni/Fe]} = 0.02\pm 0.05$, and in Luck \& Heiter (2007) -- ${\rm [Ni/Fe]} = 0.01\pm 0.03$, 
so it can be said that the [Ni/Fe] ratios in clump stars are approximately solar. 

Vanadium was investigated in our and two 
other studies. We obtain the mean value of 
${\rm [V/Fe]}=0.11\pm 0.12$, Liu et al.\ derived $0.02\pm 0.10$ and Luck \& Heiter found $-0.05\pm 0.09$,  
which means that this element also has solar [V/Fe] ratios. 
       
Chromium and cobalt were investigated in our work and by Luck \& Heiter. 
The mean [Cr/Fe] ratios are exactly solar in both studies. [Co/Fe] is enhanced by about 
$0.07\pm 0.06$~dex in the work by Luck \& Heiter. In our study, cobalt abundances 
were investigated with 
hyperfine structure effects taken into account; 
we find ${\rm [Co/Fe]}=0.02\pm 0.11$, which is close to solar.  

\subsection{Final remarks}
The main atmospheric parameters  $T_{\rm eff}$, log~$g$, $v_{t}$, [Fe/H] and abundances of vanadium, chromium, 
cobalt and nickel were determined for 62 red clump stars revealed in the Galactic field by the {\it Hipparcos} 
orbiting observatory.   

The stars form a homogeneous sample with the mean value of temperature  
$T_{\rm eff}=4750\pm 160$~K, of surface gravity log~$g=2.41\pm0.26$ and the mean value 
of metallicity ${\rm [Fe/H]}=-0.04\pm0.15$. 
It is especially interesting to note that metallicities of stars in the Galactic clump lie in quite 
a narrow interval. A Gaussian fit to the [Fe/H] distribution produces the mean 
$\langle{\rm [Fe/H]}\rangle= -0.01$ and dispersion $\sigma_{\rm [Fe/H]}=0.08$. 

The near-solar metallicity and small dispersion of $\sigma_{\rm [Fe/H]}$ of clump stars of the Galaxy obtained 
in this work confirm the theoretical model of the {\it Hipparcos} clump by Girardi \& Salaris (2001) 
which suggests that nearby clump stars are (in the mean) relatively young objects, reflecting mainly the 
near-solar metallicities developed in the local disk during the last few Gyrs of its history. 

The iron group element to iron abundance ratios in the investigated clump giants are close to solar. 
This allows us to use the clump stars to study the chemical and dynamical evolution of the Galaxy. 
Clump giants may provide a very useful information on mixing processes in evolved low mass stars. 
We plan to address this question in our further study of the Galactic clump. 
       
\section*{Acknowledgments}

This project has been supported by the European Commission through the Baltic Grid project as well as 
through ``Access to 
Research Infrastructures Action" of the ``Improving Human Potential Programme", awarded 
to the Instituto de Astrof\' isica de Canarias to fund European Astronomers' access to the 
European Nordern Observatory, in the Canary Islands. 
EP and GT acknowledge support from the Lithuanian National 
Science and Studies Foundation through GridTechno project. 
JGC acknowledges the inspiration of the late Bohdan Paczynski for stimulating her interest in this area 
and is grateful to NSF grant AST-0507219 and AST-0908139 for partial support. 
DFG is grateful for financial support from the 
Natural Sciences and Engineering Research Council of Canada. 
SJA thanks Dr. James E. Hesser, Director of the Dominion Astronomical Observatory for the observing time 
used for this project and Dr. Austin Gulliver for help in reducing the DAO Spectrograms. Financial support 
was provided to SJA by The Citadel Foundation.

\bsp

\label{lastpage}

\end{document}